\documentclass[review]{elsarticle}
\usepackage{float}
\usepackage{geometry}
\usepackage{epstopdf}
\usepackage{lineno,hyperref}
\usepackage{amsmath}
\journal{Journal of \LaTeX\ Templates}

\begin{document}

\begin{frontmatter}

\title{Effects of controlling parameter on symbolic nonlinear complexity detection}

\author[mymainaddress]{Wenpo Yao}
\author[myfirstaddress]{Min Wu}
\author[mysecondaryaddress]{Jun Wang\corref{mycorrespondingauthor}}
\cortext[mycorrespondingauthor]{Corresponding author}
\ead{wangj@njupt.edu.cn}

\address[mymainaddress]{School of Telecommunications and Information Engineering, Nanjing University of Posts and Telecommunications, Nanjing 210003, Jiangsu, China}
\address[myfirstaddress]{Jinling Hospital, Nanjing 210002, Jiangsu, China}
\address[mysecondaryaddress]{Smart Health Big Data Analysis and Location Services Engineering Lab of Jiangsu Province, Nanjing University of Posts and Telecommunications, Nanjing 210023, Jiangsu, China}

\begin{abstract}
Symbolic transformation, a coarse-graining process, is a crucial prerequisite for and has evidential influence to the symbolic time series analysis. We employ Shannon entropy for a parameter-dependent symbolization, KW (Kurths-Wessel) symbolic method, to test the effects of controlling parameter on its symbolic nonlinear complexity detection. Two chaotic models, logistic and Henon series, and heartbeats of CHF (Congestive Heart Failure) patients, healthy young and elderly subjects from PhysioNet are applied to test the KW symbolic entropy. The complexity-loss theory about aging and diseases in heart rates is validated and reasons that may account for some paradoxes in nonlinear analysis are discussed. Tests results suggest that due to different structural or dynamical properties of different nonlinear systems, controlling parameter of the KW symbolization should be adjusted accordingly to have reliable symbolic nonlinear analysis.
\end{abstract}


\end{frontmatter}


\section{Introduction}
Symbolic dynamic deals with robust properties of dynamics without digging into numbers and provides a rigorous way of looking at real dynamics with finite precision \cite{Hao1991}. And the symbolic time series analysis \cite{Daw2003}, involving transforming raw time series into symbolic sequence consisting of several discretized symbols, simplifies the complex system analysis with features of fast calculation, robustness, insensitivity to noise and so on. At the same time, due to the generally severe degree of the symbolic coarse-graining processes, some detailed information are lost and the symbolic transformations may have effects on time series analysis \cite{Xu2011}.

Due to the structural or dynamical differences in different systems, some symbolic transformations have adjustable controlling parameters to be adaptive and flexible. A symbolic transformation with adjustable parameter belonging to static methods that transform series into symbolic sequences by dividing series into finite partitions in the works of Kurths J. and Wessel N. et al. \cite{Kurths1995,Wessel2000,Wessel2007} has effective applications in complex time series analysis. It calculates the time series mean and sets a controlling parameter ¦Á to transform series into symbol sequence on the basis of the alphabet $ A=\{0,1,2,3\} $. In some follow-up researches, scholars find it is effective in symbolic nonlinear features detection of physiological signals of cardiac or brain activities \cite{Yao2017M,Wang2014,Shen2011}. In these reports, however, there is no in-depth research on the effects of controlling parameter on symbolic dynamics analysis. The parameter ¦Á determines the three partitioning lines for the formulation of symbolic sequence and will affect the nonlinear dynamics extraction. Systematical researches prove that coarse-graining symbolic methods have effects on the scaling properties of correlated and anti-correlated signals \cite{Xu2011}, then how the controlling parameter in KW (Kurths-Wessel) symbolic method affects nonlinear dynamics complexity detection is what we focus on in this paper.

In our contribution, we employed two chaotic models, logistic and Henon series, and three groups of heartbeats from PhysioNet to test the effects of the KW symbolization, particularly its controlling parameter, on nonlinear complexity detection based on Shannon entropy.

\section{Methods}

\subsection{Symbolic Shannon entropy}
Symbolic transformation that maps time series into sequence from a given alphabet is the first step of symbolic time series analysis. The symbolic coarse-graining procedure that some detailed information is lost while detecting the coarse dynamic behavior \cite{Daw2003,Kurths1995,Wessel2000}, therefore, it makes a compromise between extracting some dynamical information and maintaining sufficient statistics. A pragmatic symbolization is proposed for physiological time series in contributions of Kurths and Wessel et al. \cite{Kurths1995,Wessel2000,Wessel2007}, and it is a context-dependent transformation having feature of close connection to physiological phenomena and easy interpretation. Assuming time series $X=\{ x_{1},x_{2},\ldots,x_{L}\}$, the KW symbolization transforming $X$ into symbolic sequence $S$ is carried on as Eq.~(\ref{eq1}):
\begin{eqnarray}
\label{eq1}
s_{i}(x_{i})=
  \left\{
       \begin{array}{lr}
          0: (1+\alpha)\mu < x_{i} < \infty \\
          1:  \mu < x_{i} \leq (1+\alpha)\mu \\
          2: (1-\alpha)\mu < x_{i} \leq \mu  \\
          3: 0< x_{i} \leq (1-\alpha)\mu
       \end{array}
  \right.
\end{eqnarray}
where $\mu$ is the series mean and $\alpha$ is special controlling parameter with reference from 0.03 to 0.07, and in the original literatures analyzing heart rate nonlinear dynamics, $\alpha$ is set to 0.05 \cite{Wessel2000,Wessel2007}.

or the convenience of representation, there are negative values in some physiological signals. In some symbolic works referencing to this method, therefore, time series are divided into positive and negative parts which are symbolized separately, and the symbolization evolves into Eq.~(\ref{eq2}):
\begin{eqnarray}
\label{eq2}
s_{i}(x_{i})=
  \left\{
       \begin{array}{lr}
          0: (1+\alpha)\mu_{1} < x_{i} < \infty    \quad or \quad   -\infty < x_{i} < (1+\alpha)\mu_{2} \\
          1: \mu_{1} < x_{i} \leq (1+\alpha)\mu_{1}  \quad or \quad   (1+\alpha)\mu_{2} \leq x_{i} < \mu_{2} \\
          2: (1-\alpha)\mu_{1} < x_{i} \leq \mu_{1}   \quad or \quad   \mu_{2} \leq x_{i} < (1-\alpha)\mu_{2} \\
          3: (1-\alpha)\mu_{2} \leq x_{i} \leq (1-\alpha)\mu_{1}
       \end{array}
  \right.
\end{eqnarray}
where $\mu_{1}$ and $\mu_{2}$ are the means of positive and negative parts.

As a comparison, BS (base-scale) symbolization in Eq.~(\ref{eq3}) of Li and Ning \cite{Li2006,Li2007}, shares some similarities to the KW symbolic methods that BS method is also a 4-symbol method with $\alpha$, where $\mu$ is the mean of time series and $BS = \sqrt{\sum^{L-1}_{i=1} [x(i+1)-x(i)]^{2}/L-1}$ represent the root mean square of every two continuous data difference. To compare with the KW method, we do not consider m-dimensional phase space construction in the origination of base-scale entropy.
\begin{eqnarray}
\label{eq3}
s_{i}(x_{i})=
  \left\{
       \begin{array}{lr}
          0: x_{i} > \mu + \alpha \times BS \\
          1: \mu < x_{i} \leq \mu + \alpha \times BS \\
          2: \mu - \alpha \times BS < x_{i} \leq \mu \\
          3: x_{i} \leq \mu - \alpha \times BS
       \end{array}
  \right.
\end{eqnarray}

After symbolization, the next step is to construct m-bit symbolic sequences, leading to a maximum of $ 4^{m} $ different words. In our following complexity detection of classical chaotic models and real-world physiological signals, 3-bit encoding is applied to symbolic sequences. There are several statistical approaches that characterize symbol sequences such as direct visual histograms or quantitative measures based on classical statistics and information theory~\cite{Daw2003}. Entropy reflects the static properties and could be applied to original time series, and it is a useful measure for describing the probabilistic distribution~\cite{Xiong2017}. In our symbolic entropy, we count probability of each code $P(\pi)=\{p(\pi_{1}),p(\pi_{2}),\ldots,p(\pi_{N})\} $ and use classical Shannon entropy for nonlinear dynamic complexity extraction as Eq.~(\ref{eq4}), and its normalized form in KW and BS symbolic entropy is $ h(m)=H(m)/log4^{m} $.

\begin{equation}
\label{eq4}
H(m)=-\sum_{i} p(\pi_{i}) log p(\pi_{i}), \quad where \quad p(\pi_{i}) \neq 0
\end{equation}

\section{Chaotic models tests}

\subsection{Logistic map}
\begin{figure}[hbt]
  \centering
    \includegraphics[width=8cm,height=3.3cm]{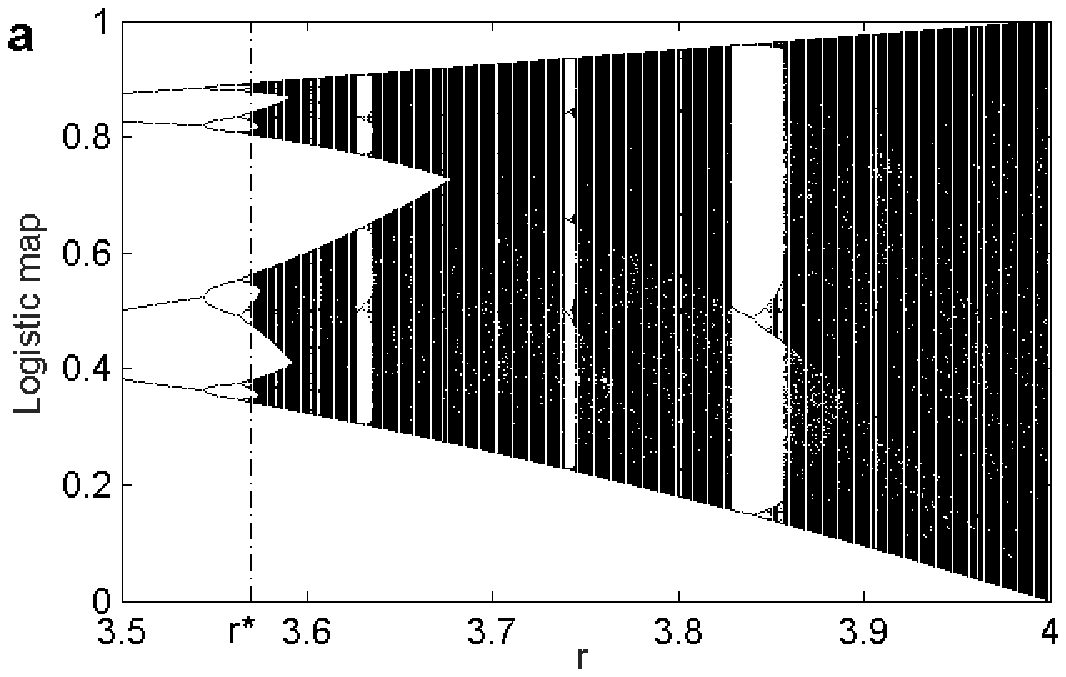}
    \includegraphics[width=8cm,height=3.3cm]{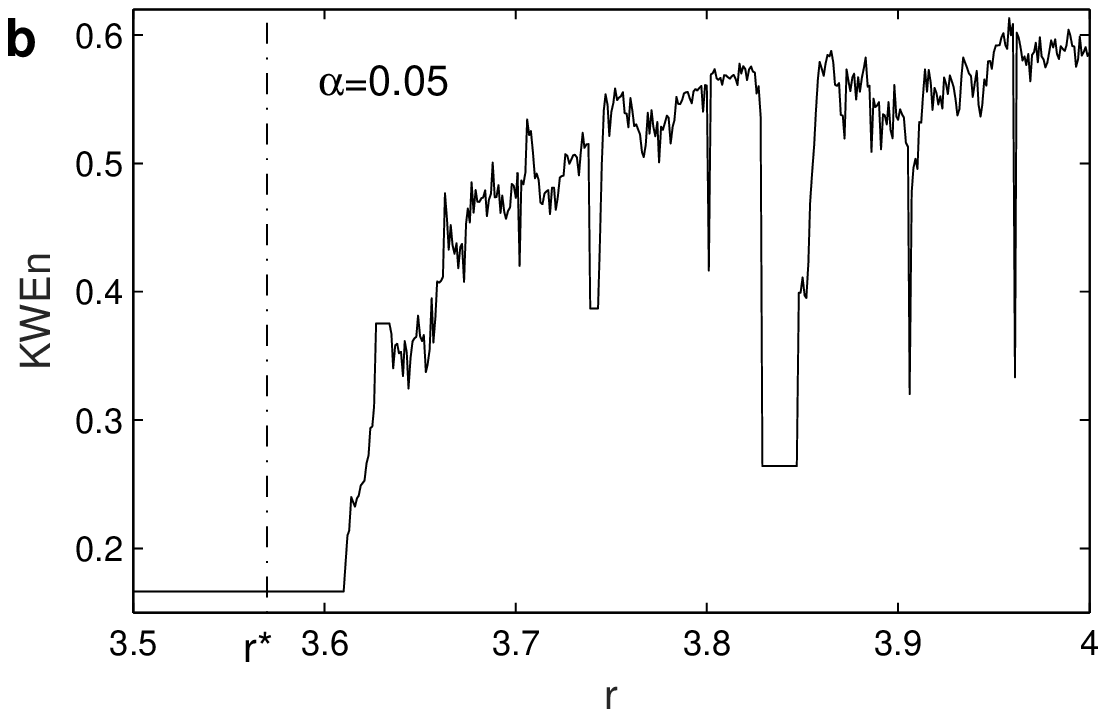}
    \includegraphics[width=8cm,height=3.3cm]{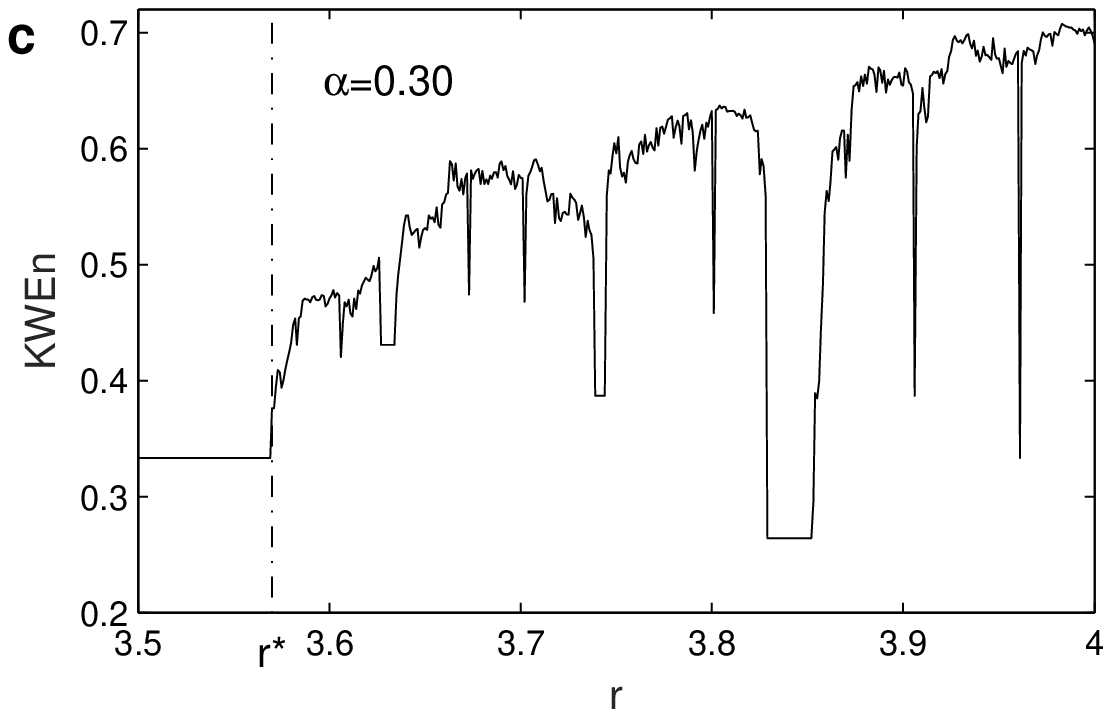}
    \includegraphics[width=8cm,height=3.3cm]{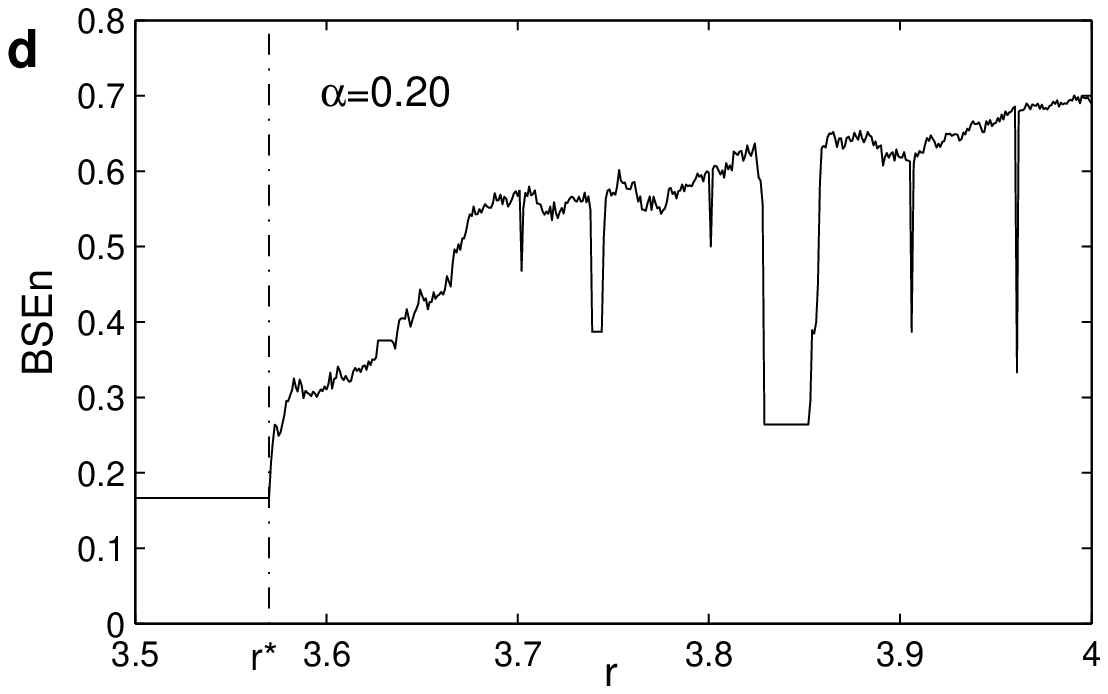}
  \caption{Logistic equations for varying control parameter from 3.5 to 4, where r*=3.567, exactly 3.569946, is the critical point for logistic series to step into chaotic sate. a) Bifurcation diagram of logistic map.  b) Lyapunov exponent.  c) KW symbolic entropy of $\alpha =0.05$. d) BS symbolic entropy of $\alpha =0.20$}
  \label{fig1}
\end{figure}

The canonical form of logistic difference equation, $ x_{n+1}=r \cdot x_{n}(1-x_{n}) $, is attractive by the virtue of its extreme simplicity \cite{May1976} and is widely applied in chaotic and nonlinear dynamical analysis. The two-degree polynomial mapping is often referenced as an archetypal example of nonlinear dynamical equations producing chaotic behaviors \cite{Boeing2016}. The logistic series (data length is 2000 and $x_{1}$=0.001) becomes chaotic at the cutting point r* shown in Fig.~\ref{fig1}a and its chaotic behaviors enhance with the increase of r.

KW symbolic entropy with the referenced controlling parameter of 0.05 does not show satisfied results between r $=$ 3.570 and 3.611 when logistic sequence becomes chaotic shown in Fig.~\ref{fig1}b. While when the controlling parameter is 0.30, the KW symbolic entropy has desirable complexity detections in the whole regions that it has accurate chaotic complexity captures as shown in Fig.~\ref{fig1}c.

Chaotic detection tests show that KW symbolic entropy with $ \alpha $ selected from the original interval cannot provide effectively complexity detections while the symbolic entropy with $ \alpha $ chosen close to 0.30 has satisfied complexity detections in the whole range of logistic map.

Similar to KW symbolic entropy, we need to adjust the controlling parameter of base-scale entropy for chaotic detection of logistic series, and we find the controlling parameter should be chosen around 0.20 shown in Fig.~\ref{fig1}d.

\subsection{Henon equations}
The Henon map is a classical two-dimensional dissipative quadratic map given by the coupled equations $x_{n+1}=1-\alpha \cdot x^{2}_{n}+y_{n}$, $y_{n+1}=\beta \cdot x_{n}$, and the dynamical system is chaotic for the classical values $\alpha=1.4$ and $\beta=0.3$ \cite{Henon1976}. We fix $\beta=0.3$ and set $\alpha$  to 1.33, 1.38 and 1.40 to observe the effects of controlling parameters on symbolic entropy in the chaotic system. Lyapunov exponents of the three Henon series (data length is 2000 and $x_{1}$=$y_{1}$=0.001) are 0.3250, 0.3702 and 0.4077.

\begin{figure}[htb]
  \centering
    \includegraphics[width=8cm,height=3.3cm]{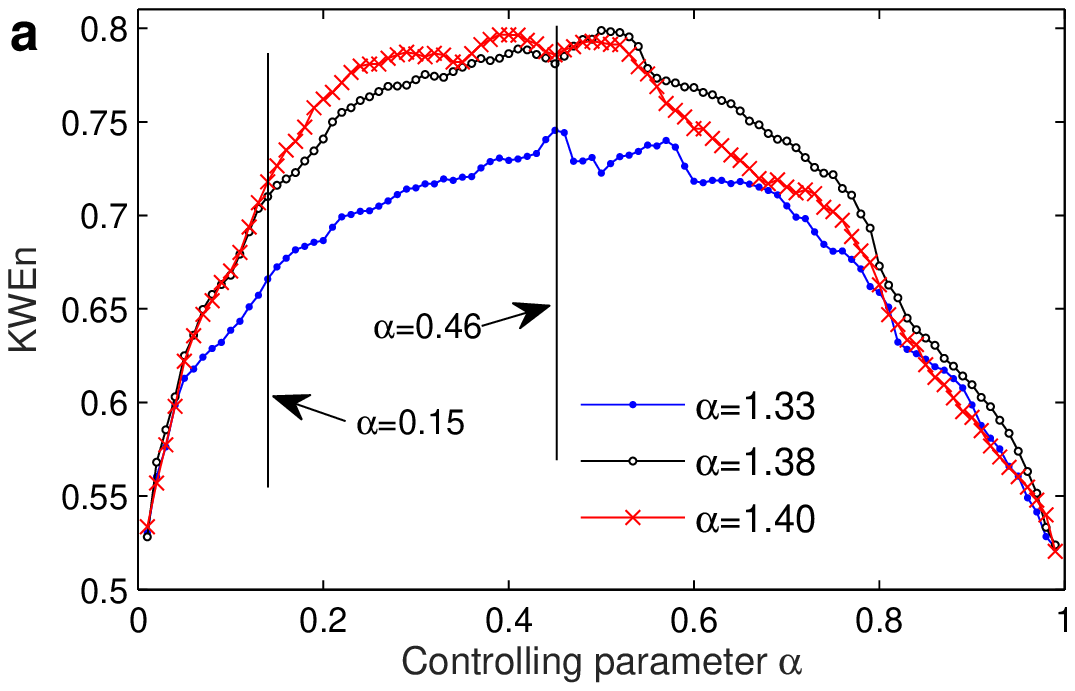}
    \includegraphics[width=8cm,height=3.3cm]{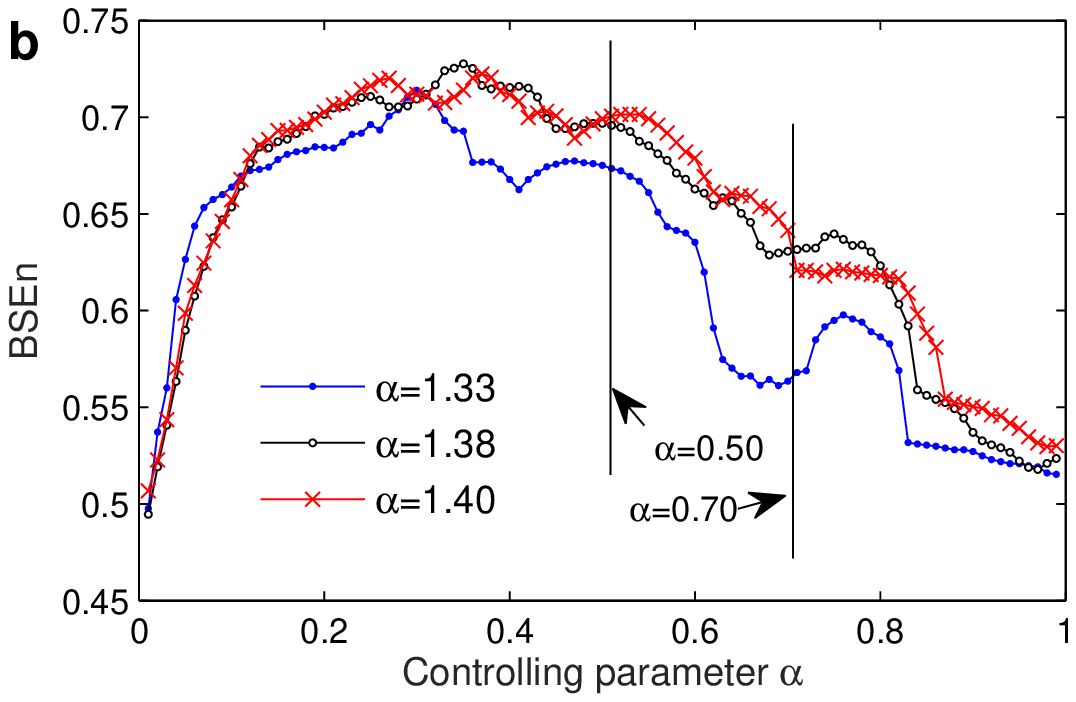}
  \caption{Symbolic entropy of three Henon outputs with fixed $ \beta=0.3 $ and $ \alpha $ = 1.33, 1.38 and 1.40. a) KW symbolic entropy. b) Base-scale entropy.}
  \label{fig3}
\end{figure}

From Fig.~\ref{fig3}, the relationships of the three Henon signals' symbolic entropy undergo a variety of changes as controlling parameter increases from 0.01 to 0.99 with step size of 0.01. Referring to Lyapunov exponents, the KW symbolic method with $\alpha$ in parameter interval [0.15, 0.45] has rational distinctions. As for base-scale entropy, their controlling parameter also need to be chosen, and intervals [0.51, 0.69] and [0.81, 0.94] have same chaotic extractions to the KW symbolic method and Lyapunov exponents.

From the nonlinear complexity detections of KW and BS symbolic entropy in the two chaotic models, we learn that due to the structural and dynamical differences between logistic and Henon systems, controlling parameter of the symbolic transformation should be adjusted differently to have reliable outcomes.

\section{Real-world physiological data tests}

\subsection{Effects of controlling parameter on symbolic dynamics extraction of heartbeats}
Three groups of heartbeats of different subjects, CHF (Congestive Heart Failure) patients \cite{Baim1986}, healthy elderly and young people \cite{Iyengar1996} from PhysioBank \cite{Goldberger2000}, are applied to test the symbolic entropy. The 15 patients (11 men aged 22 to 71 and 4 women aged 54 to 63) with severe congestive heart failure underwent about 20 hours data collection. The two healthy groups, 20 young (21 to 34 years old) and 20 elderly (68 to 85 years old) subjects, both have 2 hours of continuous supine resting while continuous data collection. Each subgroup of the healthy subjects includes equal numbers of men and women. Heart rates derived from ECG (Electrocardiography) are applied in this section.

\begin{figure}[htb]
  \centering
    \includegraphics[width=8cm,height=5cm]{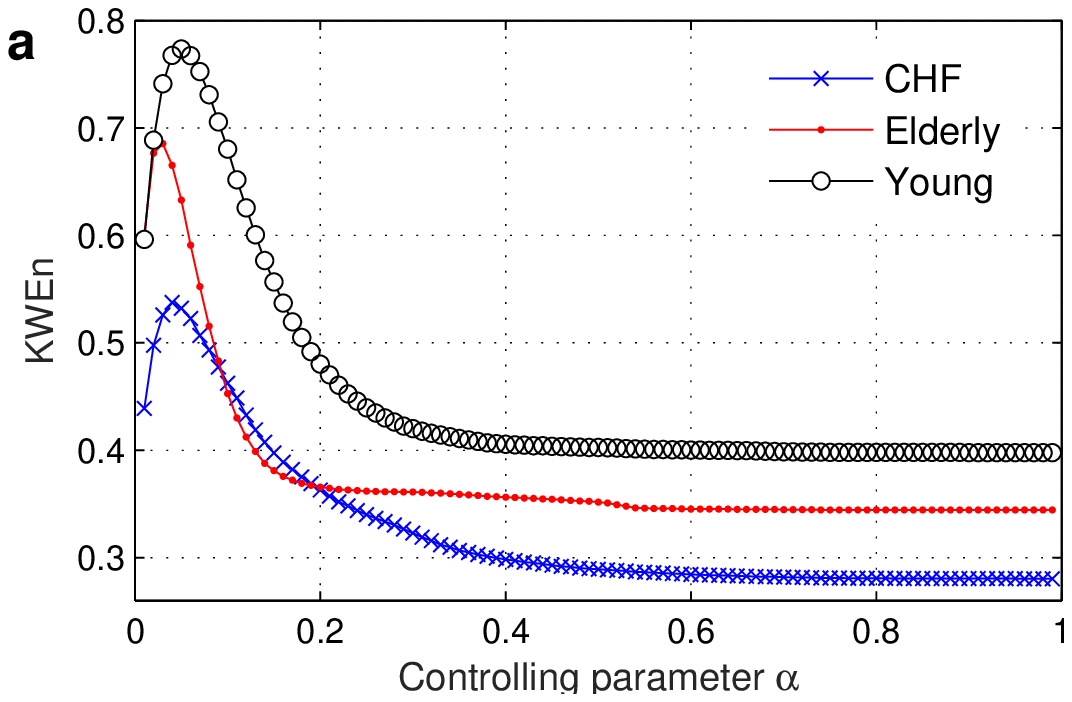}
    \includegraphics[width=8cm,height=5cm]{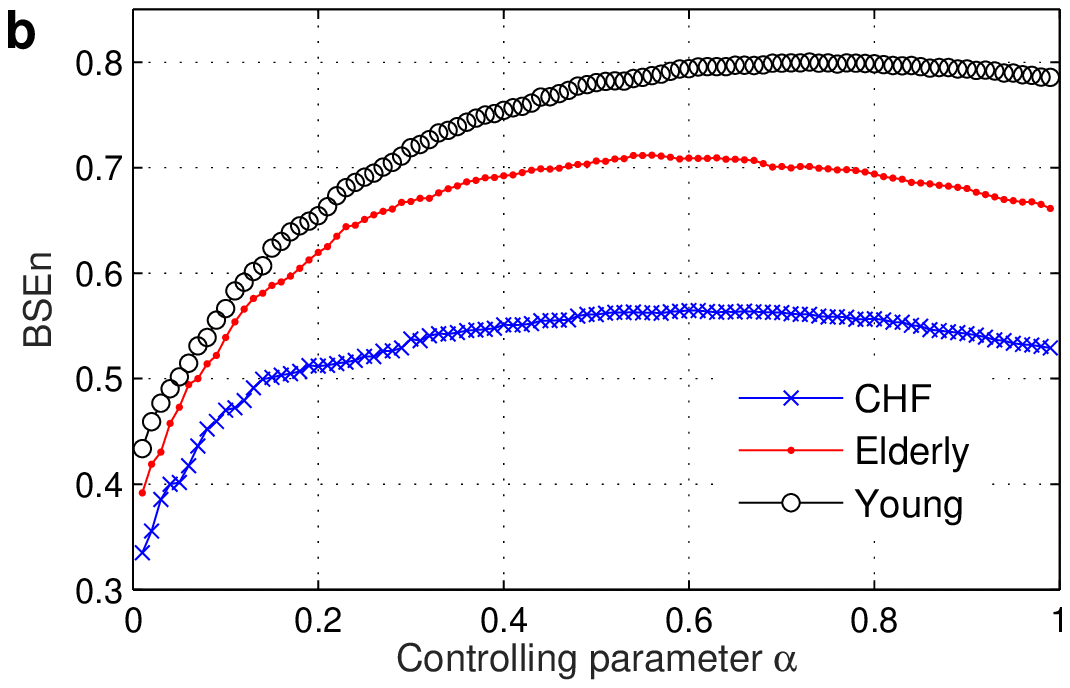}
  \caption{Two symbolic entropy methods of three kinds of heartbeats with increasing controlling parameter. a) KW symbolic entropy. b) Base-scale entropy.}
  \label{fig4}
\end{figure}

Heart rate is typical nonlinear signal with high degree of nonlinear complexity, nonlinearity and nonstationarity \cite{Wessel2000,Wessel2007,Li2006,Yao2017D,Yao2019Q}. With controlling parameter increasing from 0.01 to 0.99, the relationships of the three kinds of heartbeats' KW and BS symbolic entropy are depicted in Fig.~\ref{fig4}.

The choice of parameter is important to three kinds of heartbeats¡¯ KW nonlinear dynamic analysis indicated by Fig.~\ref{fig4}a. Healthy young volunteers¡¯ heart rates maintain higher KW symbolic entropy than the elderly ones, and CHF patients¡¯ heart signals generally have lowest nonlinear complexities. The KW symbolic entropies of three groups of subjects are divided into three parts: less than 0.05 is the first interval, from 0.06 to 0.50 is the second part, and more than 0.50 is the third one. KW entropy values of the three groups heartbeats experience rapid increases to their maximum in the first stage and undergo different degrees of decline and come to their convergences in the second section. Healthy young heart signals complexity converges to around 0.4 when ¦Á is bigger than 0.40, and the elderly subjects¡¯ symbolic entropy drop to about 0.36 when controlling parameter increases from 0.15 to 0.20 and bigger. When ¦Á is bigger than 0.50, KW entropies of the three groups of heartbeats are convergent and their relationships, the healthy young $>$ the healthy elderly $>$ the CHF, are stable.

KW symbolic entropy of three kinds of heartbeats with 6 typical parameters from different stages are chosen for statistically t tests, where 0.03, 0.04 and 0.05 are from the referenced interval in the original literatures \cite{Wessel2000,Wessel2007}, and 0.40, 0.50 and 0.60 are chosen from the convergent stage in Fig.~\ref{fig4}a. Independent samples t tests for each two heart rates¡¯ entropy values are listed in Table~\ref{tab1}.

Statistical t tests for each two heart rates' KW entropy values are displayed in Table~\ref{tab1}. P values of the CHF and healthy young are all smaller than 0.001 and are not listed in the Table.

\begin{table}[htb]
\centering
\caption{Three typical relationships of three kinds of HRV ('0.000' should be read as p $<$ 0.001). 0.03, 0.04 and 0.05 are in the referenced range of the original literatures, and 0.40, 0.50 and 0.60 are chosen from convergent stage}
\label{tab1}
\begin{tabular}{ccccc cc}
\hline
$\alpha$  &0.03	&0.04	&0.05	&0.40	&0.50	&0.60 \\
\hline
CHF-Eld 	&0.000	&0.000	&0.006	&0.036	&0.016	&0.011 \\
Eld-Yng	&0.023	&0.000	&0.000	&0.013	&0.006	&0.001 \\
\hline
\end{tabular}
\end{table}

KW symbolic entropy with the above six controlling parameters all have significant differences among the three kinds of heartbeats statistically (p $<$ 0.05), and results of parameter 0.04 and 0.05 have better distinctions. With controlling parameter in the referenced interval, however, KW symbolic entropy values of three kinds of heart signals undergo dramatic changes suggested by Fig.~\ref{fig4}a, so complexity detections are sensitive to the choice of controlling parameter. When ¦Á is bigger than 0.50, nonlinearity characterized by symbolic entropy tend to convergent, and parameters in the third stage should be better choices.

The changing trends of base-scale entropy in the three kinds of heartbeats are different from those of KW symbolic entropy, while the entropy relationships of the different kinds of heart rates are consistent with the complexity losing theory suggested by Fig.~\ref{fig4}b. According to statistical t tests to the distinctions of the heartbeats¡¯ BS symbolic entropy, when $\alpha <$ 0.18, the distinctions between of CHF and elderly are not acceptable (p $>$ 0.05), and when only $\alpha >$0.47, the separations among the three kinds of heartbeats are significant statistically.

The ¡®complexity-loss¡¯ theories of aging and disease about heart rates of related literatures \cite{Iyengar1996,Yao2017D,Yao2019Q,Ivanov1999,Goldberger2002,Costa2002,Costa2005M,Costa2008} are validated in our nonlinear complexity analysis. ECG-derived heartbeat contains valuable cardiac regulation information and has typical nonlinear dynamical behaviors which attribute to nonlinear complexity analysis for underlying mechanism of heart activities. Heartbeats of the healthy young generally yield more dynamical complexity than the elderly ones whose declined integrative cardiac control system leads to dynamical complexity reduction. Generally, it is accepted that fluctuations patterns of heartbeats in CHF patients become quite regular and nonlinear dynamics of their heart rates are severely damaged.

\subsection{Effects of data length on KW symbolic entropy}
In characterizing nonlinearity of the three groups of heartbeats, some other symbolic entropy analysis, such as the base-scale entropy \cite{Li2006,Li2007}, symbolic joint entropy \cite{Yao2017D} and so forth, have fast calculation and reliable results even with very short data sets. In the two chaotic models, we find data length does not significantly affect results, so in this subsection we conduct research on the effects of data length on the heartbeats¡¯ nonlinear complexity detection.

\begin{figure}[htb]
  \centering
    \includegraphics[width=16cm,height=9cm]{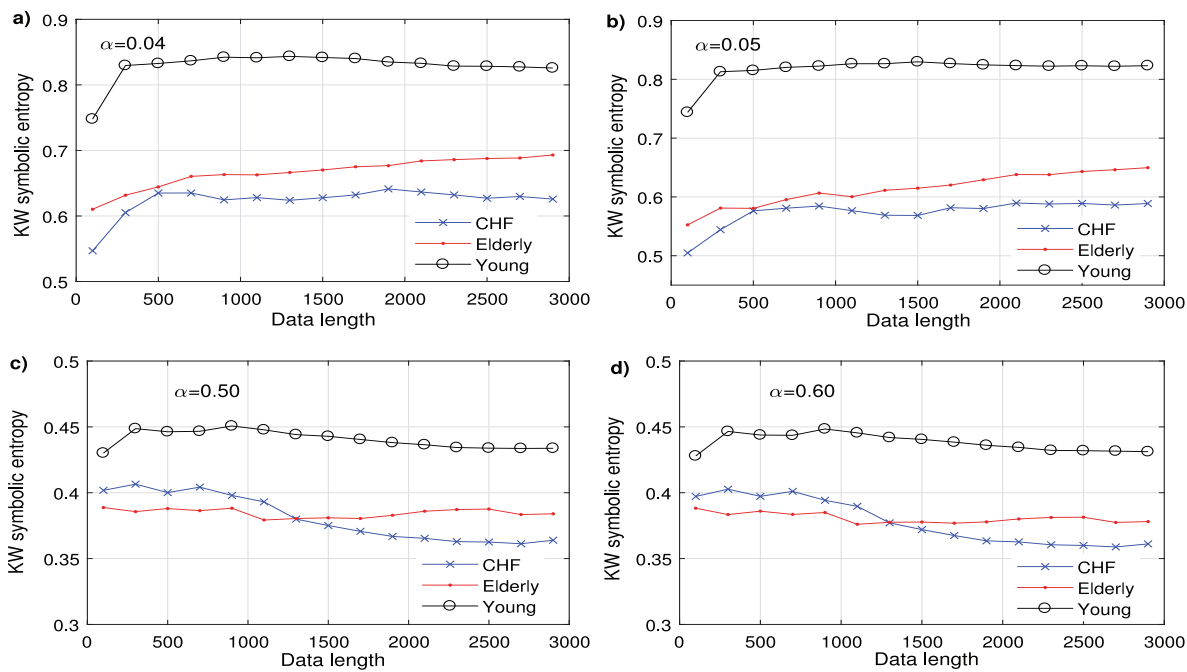}
  \caption{KW symbolic entropy of three groups of heartbeats with the increase of data length.}
  \label{fig5}
\end{figure}

KW symbolic entropy of the three different groups of subjects with data length increasing from 100 to 2900 with step size of 200 are shown in in Fig.~\ref{fig5}. We choose controlling parameters of 0.04 and 0.05 in the referenced interval in the original KW literatures \cite{Wessel2000,Wessel2007} and 0.50 and 0.60 in the convergent stage in Fig.~\ref{fig4}a.

As can be seen from Fig.~\ref{fig5}a and Fig.~\ref{fig5}b, when controlling parameters are 0.04 and 0.05, KW symbolic entropy of three groups of heart signals do not change as data length increases although distinctions between entropy of CHF heartbeats and that of healthy elderly heart rates deteriorate when data length is 500, and when the length of data sets is longer than 700, statistical differences among the three kinds of heartbeats are significant. In Fig.~\ref{fig5}c and Fig.~\ref{fig5}d, when data length is less than 1300, complexities of CHF patients¡¯ heartbeats are bigger than those of the healthy elderly which is contradictory to conventional researches, and when data length is 1700 or bigger, the three kinds of heart rates have rational complexity extraction and can be distinguished effectively. KW symbolic entropy of the healthy young subjects¡¯ heart signals maintain the highest in the four subplots in Fig.~\ref{fig5} and are not affected by data length.

From complexity extractions of the three groups of heart rates, we learn that it is more appropriate to choose ¦Á bigger than 0.50 for the stability of nonlinear analysis and insensitivity to parameter selection while controlling parameter between 0.03 to 0.05 is preferred for symbolic transformation if data length is not that large.

Our findings that the symbolic coarse-graining process has effects on symbolic time series analysis and controlling parameter should be adjusted accordingly are shared by related literatures. Xu et al. \cite{Xiong2017} demonstrated the effects of coarse-graining on the scaling behavior of time series analysis. The present authors applied symbolic joint entropy to meditation heartbeats and also found that the controlling parameters in KW and BS symbolic entropy should be adjusted to have reasonable and reliable nonlinear complexity detection \cite{Yao2018}. Also, in a symbolic approach decomposing time series into magnitude and sign series and reflecting underlying interactions in complex systems by measuring the correlations in the two sub-series by DFA \cite{Ashkenazy2001,Ashkenazy2003,Ivanov2009,Gomez2016}, the scale of DFA or scaling exponents shows effects on the distinction between the healthy and heart failure heartbeats.

\section{Discussions}
The symbolic transformation in works of Kurths J. and Wessel N. et al. is analyzed in this paper while there are still
30 some other issues that need further discussions.

\begin{figure}[htb]
  \centering
    \includegraphics[width=8cm,height=5cm]{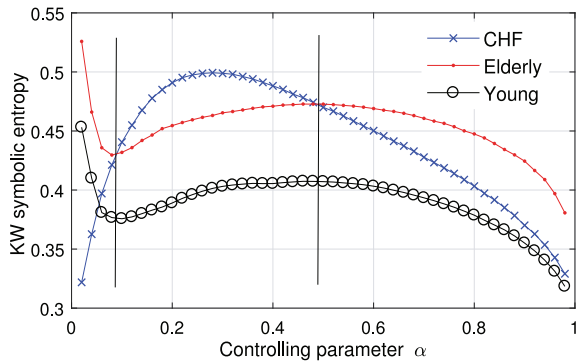}
  \caption{KW symbolic entropy of CHF, healthy elderly and young ECG.}
  \label{fig6}
\end{figure}

ECG is one of the most direct recordings of cardiac activities. Heartbeat derived from ECG is characterized with typical nonlinearity while the ECG has clear regularity due to periodic cardiac behaviors. The KW symbolic entropy for the three kinds of ECG yields some mixed results with the increase of $\alpha$ illustrated by Fig.~\ref{fig6}.

According to the Fig.~\ref{fig6}, the healthy young ECG have lower entropy values than the healthy elderly, and CHF patients¡¯ ECG symbolic entropy has significantly different changes to the two groups of healthy heartbeats groups which leads to the changes in the relationships of three kinds of ECG¡¯ KW entropy. And the results are rather confusing that symbolic entropy values of three kinds of heart signals are inconsistent with the aforementioned complexity-loss results in heartbeats. We suppose that the contradictory results are mainly due to the different nonlinear features of the heartbeat and ECG.

In other situations with perplexed nonlinear findings, we find there are some following reasons that may be responsible 9
for possible inconsistent results. 1. Multiscale factor. Multiscale analysis is to construct coarse-grained series $\{ y_{j}\}$ for the original series $\{ x_{i}\} $, $ y^{\tau}_{j}=1/\tau \Sigma ^{j\tau} _{i= (j-1)\tau +1} x_{i}$, $1 \leq j \leq N/ \tau$  where $\tau$ is scale factor. Our symbolic entropy analysis is based on single scale, however, physiologic time series like ECG may generate complexity over multiple time scale associated with a hierarchy of interacting regulatory mechanisms. The paradox maybe due to the fact that the symbolic entropy fails account for multiple time scales inherent in healthy physiologic dynamics \cite{Costa2002,Costa2005M,Costa2008}. 2. Multi-dimensional phase space. The space reconstruction goes as $Y_{m}(i)=\{ x(i),x(i+\tau),\ldots,x(i+(m-1)\tau)\}$, where $m$ is embedding dimension and $\tau$  is delay time \cite{Casdagli1991,Kim1999,Cao1997}. Low-dimensional time series, like ECG, in fact contain all information of the system, therefore some information may be hidden in higher dimensions and an approach for this situation is to construct multiple dimensional vectors to expose the hidden structural behaviors. 3. Another reason contributing to inconsistent results may be that some nonlinear features are inherent in certain forms of the system¡¯s signal and should be preproceed accordingly. The relevant nonlinear information may be reflected only by some special features extractions or preprocessing approaches. For example, the cardiac nonlinear information about physiological conditions of aging or diseases are mainly reflected by heartbeats rather than the ECG. And there might be some other reasons that are responsible for the paradox findings in nonlinear dynamics analysis remaining unknown.

In our nonlinear analysis, Shannon entropy is employed for the symbolic sequences. Entropy measures have practical and theoretical significance in applications of complexity detections of chaotic or real-world time series. In a systematic study on the entropy methods \cite{Xiong2017}, entropy measures are proved to be able to characterize the heartbeats in different physiological and clinical states and entropy is a suitable parameter in measuring the variability and preferable to be applied to the original time series. Symbolization is the first step in symbolic time series analysis, and it can be followed by a variety of different analytical methods, such as classical statistics, entropy, time irreversibility and other measures \cite{Daw2003,Yao2019Q,Yao2019TI}. Different parameters target on different aspects of information, therefore in applying other analytical measures, it need to find which controlling parameter has the optimal outcome by more comprehensive researches.

\section{Conclusions}
Through the KW symbolic entropy analysis in chaotic models and real-world heartbeats, we learn that the symbolic method should adjust its controlling parameter to achieve reliable nonlinear complexity detections because of different structural or dynamical properties of nonlinear systems. And we further verify the complexity-loss theory about aging and diseased cardiac regulation system.

\section{Acknowledgments}
The work is supported by the National Natural Science Foundation of China
(Grant Nos. 31671006, 61771251), Jiangsu Provincial Key R \& D Program (Social Development) (Grant No.BE2015700, BE2016773), Natural Science Research Major Programmer in Universities of Jiangsu Province (Grant No.16KJA310002).

\section*{References}

\bibliography{mybibfile}

\end{document}